\documentclass{article}
\usepackage[T1]{fontenc} 
\usepackage[utf8]{inputenc} 
\usepackage{ismir,amsmath,cite,url}
\usepackage{graphicx}
\usepackage{color}
\usepackage{booktabs}
\usepackage{tabularx}
\usepackage{array}
\usepackage{musicography}
\usepackage{makecell}
\usepackage{tabularray}
\usepackage{stfloats}

\title{GAPS: A Large and Diverse Classical Guitar Dataset and Benchmark Transcription Model}

\multauthor
{Xavier Riley* \hspace{1cm} Zixun Guo* \hspace{1cm} Drew Edwards \hspace{1cm} Simon Dixon} 
{Centre for Digital Music, Queen Mary University of London\\
{\tt\small j.x.riley@qmul.ac.uk, s.e.dixon@qmul.ac.uk}\\\thanks{*Equal contribution.}
}

\def\authorname{X. Riley, Z. Guo, D. Edwards and S. Dixon}

\usepackage[bookmarks=false,pdfauthor={\authorname},pdfsubject={\papersubject},hidelinks]{hyperref}

\begin{document}

\maketitle
%



\begin{abstract}
We introduce GAPS (Guitar-Aligned Performance Scores), a new dataset of classical guitar performances, and a benchmark guitar transcription model that achieves state-of-the-art performance on GuitarSet in both supervised and zero-shot settings. GAPS is the largest dataset of real guitar audio, containing 14 hours of freely available audio-score aligned pairs, recorded in diverse conditions by over 200 performers, together with high-resolution note-level MIDI alignments and performance videos. These enable us to train a state-of-the-art model for automatic transcription of solo guitar recordings which can generalise well to real world audio that is unseen during training. 

For each track in the dataset, we provide metadata of the composer and performer, giving dates, nationality, gender and links to IMSLP or Wikipedia. We also analyse guitar-specific features of the dataset, such as the distribution of fret-string combinations and alternate tunings. This dataset has applications to various MIR tasks, including automatic music transcription, score following, performance analysis, generative music modelling and the study of expressive performance timing.
\end{abstract}
\section{Introduction}\label{sec:introduction}

Automatic Music Transcription (AMT) for instruments other than piano has faced challenges due to a lack of high-quality datasets\cite{maman}. This gap has limited the development of accurate transcription systems compared to those available for the piano, which benefit from comprehensive datasets like MAESTRO\cite{maestro} and MAPS\cite{maps}. However, recent developments in audio-score alignment methods have shown promising results in improving transcription accuracy\cite{maman,hiresguitar}.

With 2.7 million guitars sold in the US alone in 2019\footnote{\url{https://www.musictrades.com/us-retail-sales-guitar-market.html}}, the guitar is a popular instrument and retains a widespread cultural significance. Around 6\% of these guitars sold were of the classical or flamenco types (roughly 162,000 units). For comparison, around 31,000 acoustic pianos were sold in the US that year.
Despite this popularity, we believe that the study of the guitar in the field of Music Information Retrieval (MIR) is underrepresented. Reviewing the paper titles for ISMIR conferences from 2013-2023 we find that publications with the word ``piano'' in the title outnumber those with ``guitar'' by 3 to 1\footnote{46 piano and 15 guitar}. This imbalance may be due to the availability of high quality datasets for piano; new datasets and methods for guitar will help to address this.

In this paper, we present GAPS, a large and diverse classical guitar dataset that contains 14 hours of matched nylon string guitar audio recordings, note-level MIDI annotations, and corresponding music scores, where the recordings feature over 200 performers in diverse recording conditions. This is several times larger than GuitarSet~\cite{guitarset}, the EGDB dataset~\cite{egdb}, the FrançoisLeduc dataset~\cite{hiresguitar} and the IDMT-SMT-Guitar dataset~\cite{IDMT-guitar} (see Section~\ref{sec:related} for a detailed comparison). We use this data to train a benchmark transcription model which achieves state-of-the-art results for solo guitar transcription across 4 dataset splits.


The contributions of this paper are as follows:
\begin{itemize}\setlength{\itemsep}{0pt}
    \item the largest available dataset consisting of real guitar audio, performance video, corresponding music scores and aligned MIDI annotations;
    \item metadata and external links for composers and performers, plus statistics of guitar-specific features;
    \item an efficient pipeline for verifying alignments of scores to audio;
    \item a benchmark state-of-the-art guitar transcription model trained on our dataset; and
    \item analysis and discussion of the effects of dataset quality, quantity and variety on AMT performance.
\end{itemize}
\section{Related Work}\label{sec:related}

\begin{table*}[t]
\centering
\resizebox{\textwidth}{!}{
\begin{tabular}{lrrrrrrrrr}\toprule
Name &Audio type &Track count &Duration (m) &Note count &Scores \\
\midrule
GuitarSet\cite{guitarset} &Real &360& 180 &62,476 &No \\
IDMT-SMT-Guitar\cite{IDMT-guitar} &Real & 1173 & 340 & $^{\ast}$5,767 & No \\
EGDB\cite{egdb} &Real &240 & 118 & 35,700 &No \\
FrançoisLeduc\cite{hiresguitar} &Real & 79 & 240 & 75,312 &Yes (commercial) \\
GAPS (ours) &Real &300 &843 &259,410 &Yes \\
\bottomrule
SynthTab\cite{synthtab} &Synthetic & 20,715 & 786,774 & - & Yes, via DadaGP \\
\bottomrule
\end{tabular}
}
\caption{Comparison of existing guitar datasets, split into real and synthetic sources. $^{\ast}$ For IDMT, the note count is shown only for notes with annotations available.}
\label{tab:dataset_comparison}
\end{table*}

GuitarSet \cite{guitarset} is the most widely used MIR dataset for guitar. It provides around 3 hours of annotated guitar performances, where the data collection process required the use of a specialised guitar fitted with a hexaphonic pickup which was able to capture the output of individual strings. The use of a single guitar severely limits the diversity of timbres and recording conditions, and in turn makes it harder for AMT models to generalise from this data\cite{hiresguitar}. 

The EGDB~\cite{egdb} dataset contains 2 hours of guitar audio recorded by a professional guitarist using a hexaphonic pickup and recorded via DI (direct input). The DI signal is then further rendered using 6 different amplifier emulation plugins. The onsets and offsets of each note are annotated. 

The IDMT-SMT-GUITAR database~\cite{IDMT-guitar} is recorded by 3 musicians using 6 different guitars (5 electric, 1 acoustic). The final audio is either obtained from DI or microphone output. It contains 4 subsets each targeting a different MIR task, ranging from single notes to chords to various short musical pieces. Its utility in transcription tasks is limited however, as only a subset of the audio has corresponding time-aligned note annotations.


Improvements in diversity of audio sources were achieved by Maman and Bermano\cite{maman} through the use of score alignment techniques. Digital scores (in MIDI format) were aligned to the activations of the Onsets and Frames transcription model \cite{oaf} trained on synthetic data. Low quality alignments were discarded and the remaining data was used to fine tune the model further. This expectation maximisation approach yielded a new state-of-the-art result on GuitarSet in the zero-shot setting, which demonstrated a generalisable model. The authors collected 5 hours of classical guitar recordings and scores in this work but these were not released as part of the publication.

Building on this approach, Riley et al.~\cite{hiresguitar} published a new state-of-the-art model for guitar transcription. Instead of the Onsets and Frames model, they use the high resolution piano transcription model by Kong et al.\ \cite{kong}, which was shown to be more tolerant of misaligned labels. Furthermore, instead of bootstrapping the process with synthetic data, they employ a pre-training step where a model is trained on the MAESTRO dataset with data augmentation, which was shown to improve generalisation. A dataset of around 4 hours of audio-MIDI pairs was published with their work, however the scores are not freely available as they were purchased from a commercial source.

As an alternative to annotating real world audio, Zang et al.\cite{synthtab} recently proposed a large scale dataset of synthesised audio from a subset of the DadaGP dataset\cite{dadagp}. When used as a pre-training step, the authors note improvements in multi-pitch estimation over 3 guitar datasets. Despite the large volume of additional training data, their note level results on the GuitarSet test split (86.1\% F1 no offset) are lower than those of several other methods which use GuitarSet alone (see \cite{hiresguitar}). This suggests that synthetic data alone is not sufficient to improve AMT systems, but a full comparison with consistent use of model architectures would be needed to establish this with certainty.

\section{Overview of Dataset}\label{sec:analysis}

\begin{figure*}

\begin{minipage}[t]{1.0\linewidth}
  \centering
  \centerline{\includegraphics[clip, width=\textwidth]{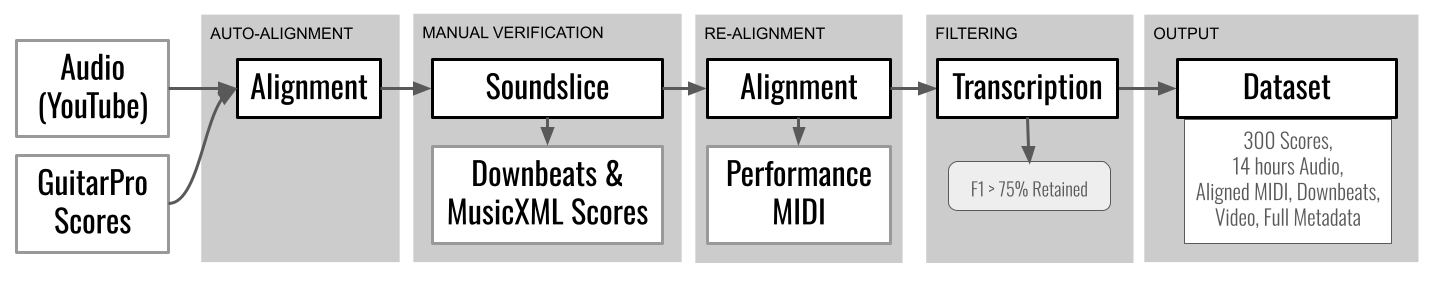}}
  \caption{Flowchart of the dataset creation process.}
  \label{fig:dataset_creation}
\end{minipage}

\end{figure*}

\subsection{Dataset Curation}\label{ssec:dataset_curation}

In an effort to improve the amount of available labelled, non-synthetic data, we have curated a new dataset of classical guitar recordings based on freely available scores from the ClassClef website\footnote{\url{classclef.com}}, together with matching performances on YouTube\footnote{\url{youtube.com}}. We align these sources using the automatic process described in \cite{hiresguitar} and then manually verified each alignment using the synchronised score viewer at \texttt{\url{soundslice.com}}. Following another alignment stage, any remaining scores with inaccurate alignments are rejected (using the criteria described below). This resulted in 300 performances sampled from the entire classical guitar canon totalling over 14 hours of music and over 250,000 note events. We have also curated extensive metadata, including information about the pieces, composers and performers, in order to enrich the dataset with details of the cultural context.

Our curation process is shown in Figure \ref{fig:dataset_creation}. It begins with the ClassClef website which provides around 5,500 pieces for download in PDF and GuitarPro formats. These focus mainly on the classical guitar with some flamenco and fingerstyle pieces included. Additionally, 547 of the pieces include links to videos on YouTube of a performance of the same piece. We first collected all GuitarPro files and converted them to MusicXML and MIDI formats using the free MuseScore software package\footnote{\url{musescore.org/en}}. We also downloaded the audio and video for the 547 pieces where YouTube links were available.

Using the alignment method described in \cite{hiresguitar}, we produce an initial alignment between the score and the recording for each piece. This proceeds in two stages: an initial alignment via Dynamic Time Warping (DTW), and a further fine alignment stage in which the notes of each chord are aligned to their closest activation from an existing transcription model. We emphasise this point as the resulting alignments are fully polyphonic in nature and as a result are more accurate than those produced by DTW alone, as described in \cite{hiresguitar}.

In some cases the automatic alignment will not succeed, for example, when a linked video contains audio for an entire suite but the score only contains a single movement. For this reason a manual verification step was required.
Using the \texttt{\url{soundslice.com}} website, we upload the automatically aligned downbeats to synchronize playback between the audio and the score. This allowed the authors of the paper (each with over 10 years of music experience) to review 474 of the scores (chosen at random) in an efficient workflow. More specifically, we manually verified the alignment between each downbeat location and the score for all 474 pieces. Particular attention was paid to the beginning and ending of each piece as these were a frequent source of issues in the DTW process. Moreover, any differences between the score and the performance that were identified were corrected, if feasible. In the end, 74 pieces were rejected for various reasons -- for example those containing 7-string guitars, guitar duets and pieces where the edition did not match the performance. Out of the remaining 400 pieces examined, 280 were usable without corrections to the score and the remaining 120 required intervention to obtain correct downbeat alignments.

The 400 reviewed scores were then re-aligned using the same alignment method from step two of figure \ref{fig:dataset_creation}. The corrected downbeats were used as anchor points during this alignment stage to ensure that any alignment errors would be localised to one measure of music. To validate accuracy, we then compared our aligned versions of the score to outputs of the guitar transcription model from \cite{hiresguitar}. We retained the 300 scores with the highest agreement, measured using the ``F-measure no offset'' metric from the \texttt{mir\_eval} library\cite{mireval}. We retained scores which had an F-measure of more than 75\%, yielding 300 audio-score pairs. We manually reviewed the lower scoring alignments and found a number of issues including errors with the processing of anacrusis bars, non-440Hz tunings and discrepancies between the performance and score editions. We hope to address these where possible as part of future work.



A summary of existing guitar datasets is shown in Table \ref{tab:dataset_comparison}. When considering datasets with real (as opposed to synthesised) audio, GAPS represents a significant advance in terms of the duration of audio and number of note events. In addition, ours is the first dataset of real audio to include freely available full music scores, tablatures in MusicXML format, and accompanying performance videos.

\subsection{Composers}\label{ssec:composers}

Works from 93 different composers are included, ranging from the Renaissance (Luys Milan, c.1500-1561) to the present day. The majority of works are from the classical guitar repertoire, with a small number of flamenco pieces and arrangements of popular music. We include the dates, nationality and presumed gender of each composer with links to canonical URLs (IMSLP and Wikipedia) where possible.

\begin{figure}

\begin{minipage}[t]{1.0\linewidth}
  \centering
  \centerline{\includegraphics[width=\textwidth]{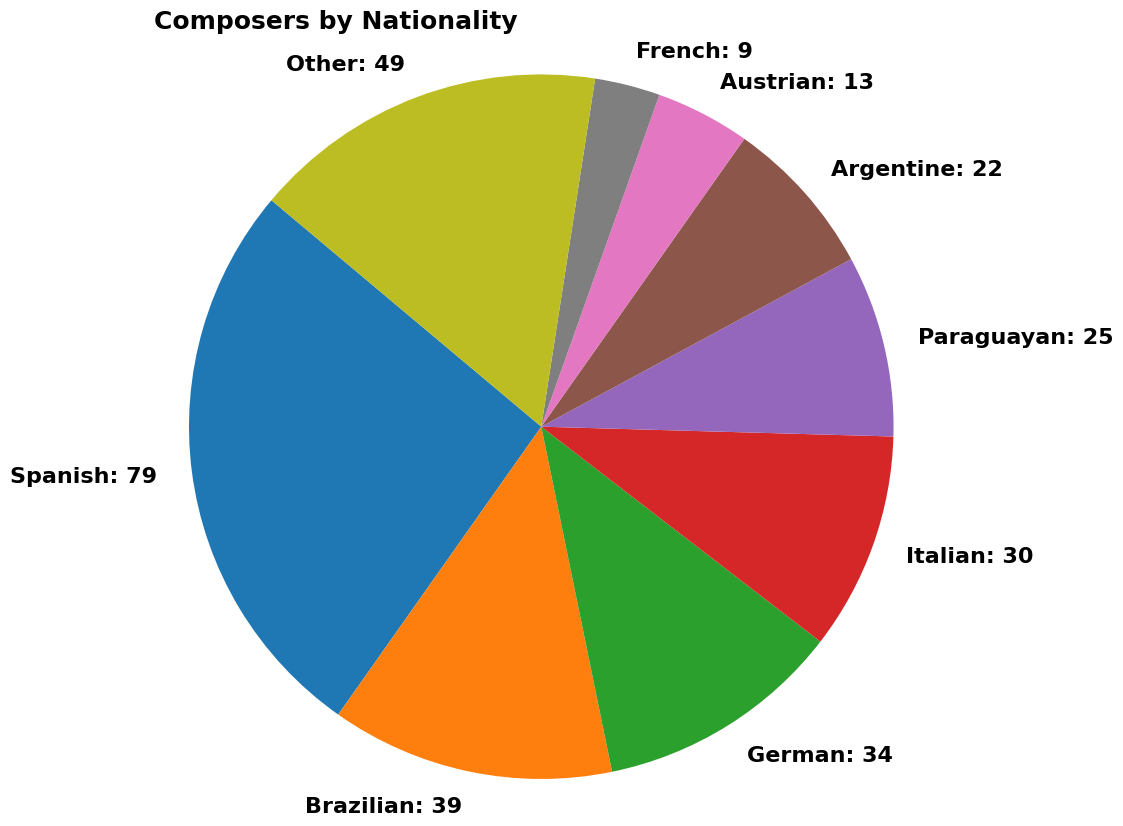}}
  \caption{Nationalities of the composers}
  \label{fig:composer_nationalities}
\end{minipage}

\end{figure}

\begin{figure}

\begin{minipage}[t]{1.0\linewidth}
  \centering
  \centerline{\includegraphics[width=\textwidth]{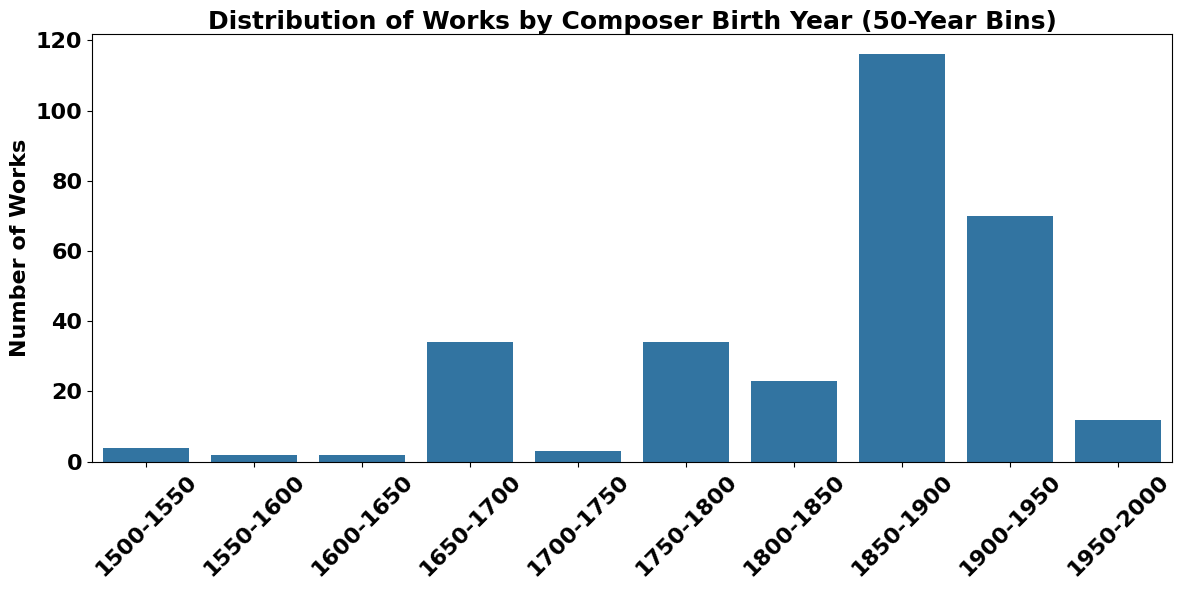}}
  \caption{Histogram of works according to composer's birth year at 50 year intervals}
  \label{fig:composer_birth}
\end{minipage}

\end{figure}

Examining the diversity of composers contained in the dataset, Figure \ref{fig:composer_nationalities} shows their nationalities, according to data from the canonical URL for each composer. This shows they are broadly divided between Europe and Latin America. In terms of chronology Figure \ref{fig:composer_birth} shows the distribution of pieces according to the year in which the composer was born. This shows that the included pieces are mainly weighted around the Romantic era (1850-1900). The peak around 1650 is almost entirely due to J.S. Bach, who is the second most common composer in our dataset with 23 pieces. We also include information about the presumed gender of composers in our metadata, however only two female composers (Maria Linnemann and Luise Walker) are included who together represent 2\% of the total by piece count. We acknowledge that this is a shortcoming of the current dataset and we will seek to address this in future work.

\subsection{Performances}

The accompanying videos are drawn from 205 different performers with YouTube views totalling over 35 million across all videos. Some are professionally produced recordings whereas others are recorded on commodity equipment such as phones and laptops. We believe this is an advantage of this dataset in that recordings are drawn from a wide variety of real world recording conditions, which in turn helps to increase the robustness of trained AMT models.

In the metadata we include information about the name of the performer (where available), their social media links (if available), the YouTube channel, the view count and the presumed gender of the performer. This was gathered to examine the extent to which classical guitar is a male dominated field. We find that female performers are better represented than composers in our dataset, but still only comprise 23\% of the total. 

\subsection{Guitar-Specific Features}

The large number of scores allows us to examine several guitar-specific features of the data. In Table \ref{tab:guitar_tunings} we see that two different tunings account for 97\% of the data. While standard tuning is most common, almost 20\% of pieces have the lowest string tuned down one tone to D.  Other alternate tunings account for around 3.3\% of the total.

\begin{table}[tb]
    \centering
    \begingroup
    \begin{tabularx}{8.2cm}{ 
   >{\raggedright\arraybackslash}X 
   >{\raggedleft\arraybackslash}X 
   >{\raggedleft\arraybackslash}X }
        \toprule
        \thead[l]{Tuning} & \thead[r]{Count} & \thead[r]{\% of total} \\
        \midrule
        EADGBE & 232 & 77.33 \\
        DADGBE & 58 & 19.33 \\
        DGDGBE & 5 & 1.67 \\
        EADF\musSharp{}BE & 2 & 0.67 \\
        FADGBE & 1 & 0.33 \\
        CGDGBE & 1 & 0.33 \\
        EBDGBE & 1 & 0.33
    \end{tabularx}
    \endgroup
    
    \caption{Distribution of guitar tunings in GAPS. The tuning is expressed from low to high pitch.}
    \label{tab:guitar_tunings}
\end{table}

\begin{figure*}

\begin{minipage}[t]{1.0\linewidth}
  \centering
  \centerline{\includegraphics[width=\textwidth]{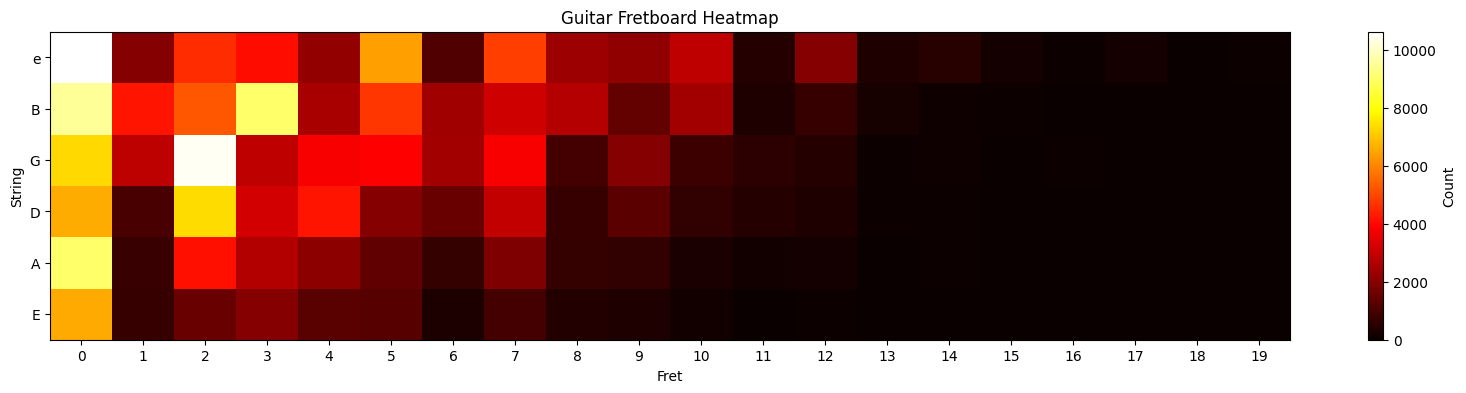}}
  \caption{Heat map of the fret/string combinations in the GAPS MusicXML tablatures.}
  \label{fig:fret_heatmap}
\end{minipage}

\end{figure*}

To see the distribution of notes across the guitar neck in this dataset, we have plotted a heat map as shown in Figure \ref{fig:fret_heatmap} using the fret information contained in the MusicXML tablature. Over the 259,000 note events we see that the pieces in the classical guitar repertoire favour the use of open strings and the first position. The strong peak at the 2nd fret A on the G string also suggests a preference towards ``guitar friendly'' keys such as E and A which allow the performer to use the open bass and top strings. While this distribution is uneven, we consider this to be representative of the classical guitar repertoire. We encourage other dataset authors to explore similar visualisations in future work to see if this varies with other genres.

Since most pitches can be played on more than one position on the guitar, there is an exponentially large number of tablatures that correspond to any one given score, including many physically unplayable versions. While each tablature in our dataset represents one valid way to play the score, we have not verified the extent to which the tablatures correspond to the choices of the performers in the specific performances in the GAPS dataset. This is left for future work. As we were not able to trace the provenance of the ClassClef data, we presume the data is crowdsourced and reflects the playing habits of a subset of computer-literate guitarists. It is also possible that some of the tabs were generated algorithmically from the score data.


\section{Transcription Baseline}
\subsection{Experimental Settings}\label{ssec:transcription_model}
To demonstrate the utility of the GAPS dataset of aligned score-audio pairs, we trained several guitar transcription models using the high resolution model of Kong et al.\ \cite{kong}, which achieved state-of-the-art performance when trained for guitar transcription \cite{hiresguitar}. This model is a convolutional recurrent neural network (CRNN) that is trained in a supervised manner to map log mel-spectrograms of 10-second segments of audio to MIDI. The convolutional layers span only across the frequency dimension, maintaining the time-resolution of the original spectrogram (10ms). These features are then processed by a gated recurrent unit (GRU) to produce the final outputs of onset, offset, frame activity, and velocity activations per pitch per time window.

There are two reasons why we used the high resolution model~\cite{kong}. Firstly to ensure fair comparisons with the state-of-the-art model in \cite{hiresguitar} as it shares the same architecture. This allows us to examine how our GAPS dataset influences the same transcription model. Secondly, fine-tuning becomes feasible due to the shared architecture among multiple piano transcription models~\cite{kong,edwards_aug}. This allows us to investigate whether different pre-trained piano transcription models can improve guitar transcription through domain adaptation.

For our experiments, we trained 2 sets of models. The first set of models is trained only on the GAPS dataset and the second set of models is trained with a combination of GuitarSet and GAPS. We employ the first set of models for zero-shot inference on the complete GuitarSet, while the second set is utilised to evaluate guitar transcription performance across the test splits of GuitarSet, theFrançoisLeduc dataset and GAPS. To study the effects of pre-training and finetuning~\cite{edwards_aug,synthtab}, each set of models has 3 variants: one trained from scratch and two finetuned from one of two published checkpoints for piano transcription \cite{kong,edwards_aug}. This also allows for a more direct comparison with results reported in \cite{hiresguitar}.

Regarding our training data and strategy, we randomly divide the GAPS dataset with a 90:10 split by piece, for training and testing respectively. Following \cite{hiresguitar}, each audio file is split into 10-second chunks, using a hop size of 1 second. We adopt the same train-test split from~\cite{hiresguitar, tfperceiver} for GuitarSet. During training, pitch shifting of up to $\pm 3$ semitones was randomly applied as data augmentation \cite{tfperceiver}.

\subsection{Transcription Results}\label{sec:results}

In Tables~\ref{tab:guitarset_sup} to~\ref{tab:gaps_test}, we report the evaluation results for the models described in Section \ref{ssec:transcription_model}. Our proposed combination of model, pre-training checkpoint and dataset achieves state-of-the-art performances on all 4 test sets mentioned in Section \ref{ssec:transcription_model}. Considering the similarities to the approach used by Riley et al.\cite{hiresguitar}, our larger dataset appears to drive the improvement in results.

\begin{table}[!b]
 \begin{center}
 \begin{tabular}{|l|l|l|l|}
  \hline
  & $P_{50}$ & $R_{50}$ & $F_{50}$ \\
  \hline
  Basic Pitch \cite{bittner} & - & - & 79.0 \\
  MT3 \cite{mt3} & - & - & 90.0 \\
  Zang et al~\cite{synthtab} & - & - & 84.5 \\
  Lu et al. \cite{tfperceiver} & - & - & 91.1 \\
  SpecTNT (in \cite{tfperceiver}) & - & - & 90.7 \\
  Riley et al. \cite{hiresguitar} ($_\textit{FL}$) & 87.6 & 86.8 & 86.9 \\
  Riley et al. ($_\textit{GS+FL}$) & 91.1 & 88.5 & 89.7 \\
    \hline
    Ours & & & \\
    \hline
  ($_\textit{GAPS}$) &89.9&85.4&87.2\\
  ($_\textit{GAPS}$ Finetuned from \cite{kong}) & 88.8&86.8&87.5\\
  ($_\textit{GAPS}$ Finetuned from \cite{edwards_aug}) & 90.1 & 86.6 & 88.0 \\ 
  ($_\textit{GAPS+GS}$) & 90.2 & 90.9&90.4\\
  ($_\textit{GAPS+GS}$ Finetuned from \cite{kong})&89.4&\textbf{92.1}&90.7\\
  ($_\textit{GAPS+GS}$) Finetuned from \cite{edwards_aug}) & \textbf{91.3} & 90.7 & \textbf{91.2} \\
  
  \hline

 \end{tabular}
\end{center}
 \caption{Results for note-level transcription accuracy on the GuitarSet test split. $P_{50}$, $R_{50}$, and $F_{50}$ are Precision, Recall and F1-measure, expressed as percentages, at 50ms resolution. All are evaluated on onsets only (no offsets or velocity), using the \texttt{mir\_eval} library. Baseline results are described in \cite{hiresguitar}.}
 \label{tab:guitarset_sup}
\end{table}

\begin{table}[t]
 \begin{center}
 \begin{tabular}{|l|l|l|l|}
  \hline
  & $P_{50}$ & $R_{50}$ & $F_{50}$ \\
  \hline
  MT3 \cite{mt3} & - & - & 32.0 \\
  Kong et al.~\cite{kong} & 67.5 & 49.7 & 54.8 \\
  Kong et al.\ (w/ aug) & 80.6 & 44.0 & 50.3 \\
  Zang et al.~\cite{synthtab}\ (Synthtab) & - & - & 70.2 \\
  Maman (MusicNet$_\textit{EM}$) \cite{maman} & 86.6 & 80.4 & 82.9 \\
  Maman (Guitar) \cite{maman} & 86.7 & 79.7 & 82.2 \\
  Riley et al. \cite{hiresguitar} & 88.0 & 87.1 & 87.3 \\
  \hline
   Ours &\textbf{92.4}&81.8&86.1\\ 
  Ours (Finetuned from \cite{kong}) & 91.6&83.7&87.0\\ 
  Ours (Finetuned from \cite{edwards_aug}) & 91.1&\textbf{85.9}& \textbf{88.1}\\ 
  \hline
  
 \end{tabular}
\end{center}
 \caption{Results for note-level transcription accuracy on the entire GuitarSet in the zero-shot setting.}
 \label{tab:guitarset_zs}
\end{table}

\begin{table}[t]
 \begin{center}
 \begin{tabular}{|l|l|l|l|}
  \hline
   & $P_{50}$ & $R_{50}$ & $F_{50}$\\
  \hline
  Basic Pitch \cite{bittner} & 54.6 & 85.0 & 66.1 \\
  Omnizart \cite{omnizart} & 63.0 & 72.1 & 67.1 \\
  MT3 \cite{mt3} & 48.8 & 57.0 & 52.4 \\
  Lu et al. \cite{tfperceiver} & 83.6 & 77.3 & 80.0 \\
  Riley et al. \cite{hiresguitar} & 83.9 & \textbf{85.5} & 84.7\\
  Ours (Finetuned from \cite{edwards_aug}) & \textbf{85.5} & 84.2 & \textbf{84.8}\\
  \hline
 \end{tabular}
\end{center}
 \caption{Results for note-level transcription accuracy on the test split of the FrançoisLeduc dataset \cite{hiresguitar}.}
\label{tab:francoisleduc}
\end{table}

\begin{table}[t]
 \begin{center}
 \begin{tabular}{|l|l|l|l|}
  \hline
   & $P_{50}$ & $R_{50}$ & $F_{50}$\\
  \hline

  
  
  Riley et al. \cite{hiresguitar} & 92.9 & 91.4 & 92.1\\
  \hline
   Ours & 94.9 & 92.1 &93.4 \\ 
  Ours (Finetuned from \cite{kong}) &94.6 &93.4 &94.0\\ 
  Ours (Finetuned from \cite{edwards_aug}) & \textbf{95.0} & \textbf{93.6}& \textbf{94.3}
 \\  

  \hline
 \end{tabular}
\end{center}
 \caption{Results for note-level transcription accuracy on a test split of the GAPS dataset.}
\label{tab:gaps_test}
\end{table}


\subsubsection{Generalisation and Guitar Types}\label{ssec:generalization}

GuitarSet contains audio for one acoustic steel string guitar recorded via microphone and also via the guitar pickup (the ``DI'' outputs). Despite our GAPS data containing only performances on nylon-stringed classical guitars, our model is able to generalise well to GuitarSet in the zero-shot setting (F-measure 88.1\% - see Table \ref{tab:guitarset_zs}). This result is interesting as it appears that timbral differences between guitars are not a strong factor in the success of the model for this task. On the other hand, GAPS does include a large range of guitars and recording conditions (unlike GuitarSet's one guitar), which we expect would contribute to the generalisation performance of models trained on it.

We also note that for the other solutions based on encoder-decoder architectures~\cite{mt3,tfperceiver}, the strong results in the supervised setting on GuitarSet fail to perform as well on unseen data. Table \ref{tab:guitarset_zs} shows the transcription accuracy on GuitarSet in the zero-shot setting, i.e.\ where models are trained without any access to GuitarSet.  F-measure scores for MT3 fall from 90.0\% to 32.0\% on GuitarSet. The previous state-of-the-art model (Time-Frequency Perceiver) from Lu et al.\ \cite{tfperceiver} attains 91.1\% in the GuitarSet supervised task but drops to 80.0\% on the unseen FrançoisLeduc test set. It may be the case that these architectures require more data to generalise effectively and we hope to explore training them on GAPS in future work.

For the FrançoisLeduc test split in Table \ref{tab:francoisleduc}, our proposed model outperforms Riley et al.\ \cite{hiresguitar} by a small margin, however their model was trained in a supervised fashion whereas this dataset was unseen by our model.

Conversely, our proposed method outperforms Riley et al.\ \cite{hiresguitar} on the GAPS test split by a margin of 2.2\% (see Table \ref{tab:gaps_test}). This indicates that, despite our method's strong generalisation (see Table \ref{tab:guitarset_zs}), it is somewhat specialised to classical guitar timbres and that the strongest results in the future may rely on the use of specific training data.

\subsubsection{Effects of Pre-training}\label{ssec:pre_training}

In each of our evaluations, we see a consistent trend whereby the model with no pre-training is surpassed by the model pre-trained on piano (MAESTRO) and fine-tuned on GAPS, which in turn is surpassed by the model pre-trained on an augmented version of MAESTRO~\cite{edwards_aug} before fine-tuning on GAPS. This illustrates the importance of pre-training and fine-tuning,  as well as data augmentation as important drivers of success in the transcription task (see Edwards et al. \cite{edwards_aug} for a detailed analysis of the effect of data augmentation on transcription generalisation).

We also note that strong results for other methods on the GuitarSet test split are obtained from models trained with a mixture of datasets~\cite{tfperceiver,hiresguitar,mt3}. One exception is Zang et al.\cite{synthtab}, who use a large corpus of synthetically rendered guitar samples for pre-training. This does not perform as well as other methods but their results were obtained from a model (TabCNN) designed for guitar tablature prediction, as opposed to a state-of-the-art transcription model. A full comparison of synthetic and real audio for pre-training is something we also hope to explore in future work.

\section{Conclusion}\label{sec:conclusion}

We present GAPS, a large dataset of score and audio pairs for solo classical guitar which comprises a wide range of composers, performers and real-world recording conditions, totaling 14 hours of recordings. The MIDI annotations are made freely available and the audio is available at the YouTube links provided. This represents the largest dataset of freely available guitar audio-score pairs to date.

We included analysis of the overall statistics of the GAPS dataset, but further musicological work could be done to examine connections between the composers, performers and musical features. The published MIDI annotations could be useful for generative modelling of classical guitar and other instruments. For future work we will look to expand the dataset and enhance the diversity where possible, particularly for the range of composers we include.

One application of this dataset is AMT for guitar, which we demonstrate through a comprehensive evaluation of a transcription model trained on our data. This shows state-of-the-art results when compared with existing methods trained on other datasets. In future work we look to examine further issues around pre-training for guitar transcription.


\section{Ethics Statement}

In addition to our role as researchers, we are also members of the global community of musicians and we seek to respect their important role in our culture. Our work here raises several issues which may have wider impact on this community which we hope to address as follows.

Firstly, we believe that using sources which are publicly available (subject to licence conditions) is important to reduce barriers to future research. At the time of writing, neither the scores nor their audio recordings are behind any kind of paywall. We have processed this data and make the results available on the basis of fostering research. We also obtained permission from the website owner of \texttt{classclef.com} to make use of their materials.

By publishing work on YouTube, artists do grant some kind of implicit licence that the data can be viewed, however the specific terms of the licence may restrict further use cases. We believe that our work is justified in using this data under fair use or fair dealing exemptions defined for research, but we are mindful that further use of the data may require express permission from the performers, composers or copyright-holders. We have attempted to address this by including detailed information about all performers and composers in the accompanying metadata to allow interested parties to contact them directly.

Finally we recognise that AMT models which approach human-level accuracy might pose a threat to those who are employed in music transcription and related fields. On the other hand, such models could also assist such work and become tools for improving the efficiency and accuracy of their daily work. For this reason we are carefully considering whether to make our model weights freely available. 

\section{Acknowledgments}
Authors XR, ZG and DE are research students at the UKRI Centre for Doctoral Training in Artificial Intelligence and Music, supported by UK Research and Innovation [grant number EP/S022694/1] and Yamaha Corporation (DE).

\bibliography{ISMIRtemplate}

%
%
%
%
%

\end{document}